\documentclass[a4paper,showpacs,citeautoscript,pra,preprint,endfloats*]{revtex4}
\usepackage[latin1]{inputenc}
\usepackage{amsmath}
\usepackage{amsfonts}
\usepackage{amssymb}
\usepackage{graphicx,color}
\begin{document}

\title{Stochastic numerical simulations of a fully spatio-temporal Hong-Ou-Mandel dip}

\author{Fabrice Devaux$^{1*}$, Alexis Mosset$^1$ and Eric Lantz$^1$}
\affiliation{$^1$ Institut FEMTO-ST, D\'epartement d'Optique P. M. Duffieux, UMR 6174 CNRS \\ Universit\'e Bourgogne Franche-Comt\'e, 15b Avenue des Montboucons, 25030 Besan\c{c}on - France}

\date{\today}
\email[Corresponding author:]{fabrice.devaux@univ-fcomte.fr}

\begin{abstract}
We develop a fully spatio-temporal numerical model, based on stochastic simulations, simulating the generation of spatio-temporal multi-mode spontaneous parametric down conversion, the propagation of the signal and idler beams through a Hong-Ou-Mandel interferometer and the detection of the outgoing beams with two separate detectors arrays. Spatial and temporal properties of the two-photon interference are investigated by measuring the spatial distribution of the momenta coincidence counts between the time integrated outgoing far-field images and a full spatio-temporal HOM dip is exhibited. Numerical results also demonstrate that the spatio-temporal coherence properties of bi-photon wave packets can be fully characterized with detectors arrays with no temporal resolution.  
\end{abstract}
\maketitle
\section{Introduction}
Spatial entanglement of photon pairs in images offers new opportunities to develop protocols for communication and parallel treatment of quantum information of potentially very high dimensionality. Although entangled photon pairs of high Schmidt number are easily produced by optical Spontaneous Parametric Down-Conversion (SPDC), the manipulation and the detection of images with quantum features is tricky. Fortunately, detector arrays with high sensitivity and high quantum efficiency, such as electron multiplying charge coupled devices (EMCCDs) or intensified charge coupled devices (iCCDs), have been available since the early 2000s and are now widely used for quantum imaging experiments \cite{moreau_imaging_2019} like demonstration of Einstein-Podolsky-Rosen (EPR) paradox in twin images \cite{moreau_einstein-podolsky-rosen_2014,lantz_einstein-podolsky-rosen_2015}, ghost imaging \cite{morris_imaging_2015,denis_temporal_2017}, quantum adaptive optics \cite{defienne_adaptive_2018}, quantum holography \cite{devaux_quantum_2019}, sub-shot-noise imaging \cite{brida_experimental_2010,toninelli_sub-shot-noise_2017} and quantum imaging with undetected photons \cite{lemos_quantum_2014}.

Among the whole experiments using entangled pairs of photons, the famous experiment of two-photon interference known now as Hong-Ou-Mandel (HOM) interference \cite{hong_measurement_1987}, is probably one of the most fascinating. This groundbreaking experiment paved the way for a multitude of experiments showing the richness of the quantum properties of SPDC and their application to original communication protocols \cite{simon_quantum_2017}. Most of these experiments and protocols use the coherence time property of the bi-photon state and measurement of the quantum properties are performed by means of bucket detectors and coincidence counters. Recently, Jachura et al. \cite{jachura_shot-by-shot_2015} extended the applications of the camera systems to the observation of HOM interference with an intensified scientific complementary metal-oxide-semiconductor (sCMOS) camera. Nevertheless, it is still the properties of temporal coincidences that are studied.

Only few studies report the spatial coherence of two-photon state by measuring the coincidence counts rate through a HOM interferometer as a function of the relative transverse displacement or rotation of one SPDC beam \cite{lee_spatial_2006,kim_spatial_2006}. However, spatial coherence is demonstrated by measures of temporal coincidence with bucket detectors and coincidence circuits. 

In this paper, using numerical simulations with realistic parameters, we will show how a HOM interferometer with a two cameras setup, used to detect purely spatial coincidence counts between the two outgoing images of the interferometer, fully resolves the spatio-temporal coherence properties of entangled photons pairs of high dimensionality.

\section{Numerical model and parameters of the simulation}
\label{sec:1}
In previous works \cite{lantz_spatial_2004,devaux_quantum_2019}, we used stochastic simulations based on the Wigner formalism which reproduce, when repeated several thousand times and averaged, all specific quantum features of SPDC. With the current model, two independent stochastic input fields with the appropriate phase-probability distribution in space and time, corresponding to the signal and idler vacuum fields in the Wigner representation (i.e. Gaussian white noise with zero mean and a random phase), are generated. The propagation and the nonlinear interaction, in a thin type 2 $BBO$ nonlinear crystal, of these stochastic fields and of the Gaussian pump pulse are then calculated by integrating the classical nonlinear propagation equations \cite{saleh_nonlinear_2001}, which are solved with a split-step algorithm. We emphasize that all features involved in the three-wave-mixing interaction along the crystal length such as chromatic dispersion, diffraction, walk-off and phase-mismatch, are considered.

\begin{figure}
\centering
\includegraphics[width=15cm]{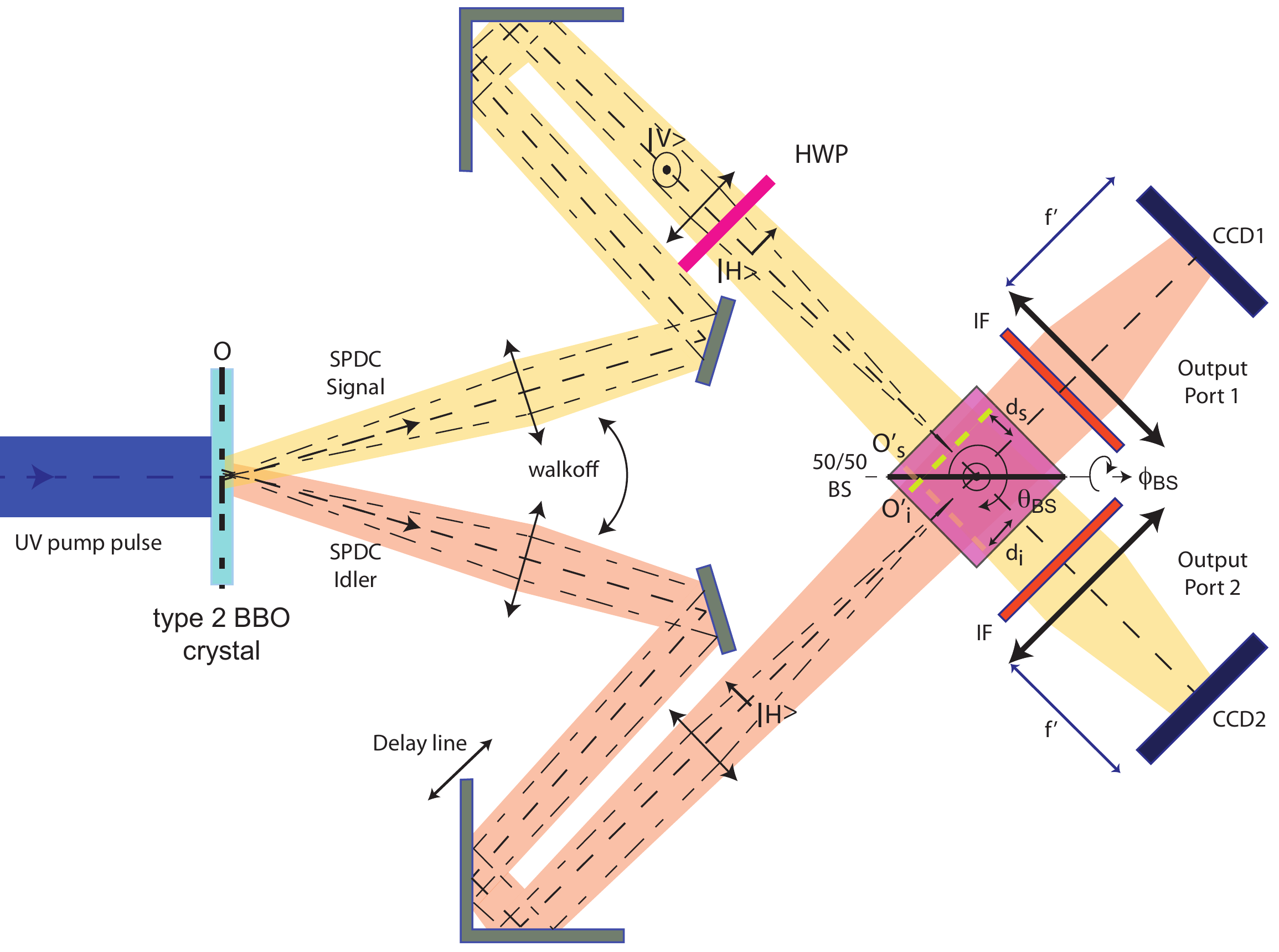}
\caption{Diagram of the HOM interferometer : Strongly multi-mode signal and idler beams generated in a type 2 BBO crystal propagate up to the two input ports of a 50/50 beamsplitter (BS). Images $O'_s$ and $O'_i$ of the near field of the crystal $O$ are made at the BS interface. Far-field of the outgoing images are detected with two separate detectors arrays. Narrow interferential filters (IF) limit the spectral bandwidth detection. Some features like optical path delay ($\delta t$), images defocusing ($d_s$, $d_i$) and transverse spatial phase shift between the transmitted and reflected beams ($\theta_{BS}$, $\phi_{BS}$) are considered within the numerical model. $\rvert H\rangle$ and $\rvert V\rangle$ symbolize the horizontal and vertical polarization states of the signal beam and of the idler beam before and after the half-wave plate (HWP).}\label{fig1}
\end{figure}

Then, propagation of the output fields through the two arms of a HOM interferometer, depicted in Fig. \ref{fig1}, is performed. From the output face of the nonlinear crystal, the signal and idler fields are propagated up to the two input ports of a 50/50 beamsplitter (BS). The near-field plane of the crystal $O$ is imaged through the two arms of the interferometer such as the transmitted and the reflected images are conjugated through semi-transparent interface of the BS. Possible defocusing distances $d_s$ and $d_i$ of the image planes $O'_s$ and $O'_i$ are considered. Time integrated far-field intensities of the outgoing fields at the two output ports of the BS are calculated to simulate the image detection on two independent detectors arrays with no temporal resolution placed in the focal planes of two identical lenses. Spectral filtering with narrow interferential filters (IF), limiting the bandwidth detection around the degenerated wavelength ($\lambda_s=\lambda_i=2\lambda_p$), is included in the model. 2D transverse spatial phase shift between the transmitted and reflected beams (induced by the rotations $\theta_{BS}$ and $\phi_{BS}$ of the BS) is also considered.   
The symbols $\rvert H\rangle$ and $\rvert V\rangle$ denote the polarization state of the two beams at different locations on the device and more particularly before and after the half-wave plate (HWP). 

Finally, these calculations are repeated a thousand times. The spatial distribution of the momentum coincidence counts between the two far-field images is obtained by calculating the normalized cross-correlation of image pairs, after subtraction from these images of their deterministic part (i.e. the mean of the images)\cite{moreau_einstein-podolsky-rosen_2014,lantz_einstein-podolsky-rosen_2015,denis_temporal_2017,devaux_quantum_2019}. 

Thanks to the numerical model, the influence on the coincidence counts of several parameters of the interferometer, like an optical path delay between the two input ports of the BS, a spatial frequency shift along vertical and horizontal axes of the reflected beams and a defocusing of the signal and idler images are investigated. The realistic numerical values of the model parameters are gathered in table \ref{table1}.

\begin{table}
  \centering
\begin{tabular}{|c|c|c|}
\hline
  Parameter & Symbol & Value (unit)\\\hline
  Pump duration & $\sigma_{t_p}$ & 42 ($ps$) \\\hline
  Pump size & $\sigma_{x_p}=\sigma_{y_p}$ & 0.1 ($mm$)\\\hline
  Pump wavelength & $\lambda_{p}$ & 354.7 ($nm$)\\\hline
  Amplification gain & $g$ & 4.2 ($mm^{-1}$)\\\hline
  Crystal length & $L_C$ & 0.8 ($mm$)\\ \hline
  Crystal width & $l_C$ & 1 ($mm$)\\ \hline
  Sampling grid & $n_x\times n_y\times n_t $ & $128\times 128\times 128 $\\ \hline
  Spatial sampling step & $dx=dy$ & $7.8\times10^{-3}$ ($mm$) \\ \hline
  Temporal sampling step & $dt$ & $2.3$ ($ps$) \\ \hline
  IF central wavelength & $\lambda_{FI}$ & $709.4$ ($nm$) \\ \hline
  IF width & $\sigma_{\lambda_{FI}}$ & $0.2$ ($nm$) \\ \hline
\end{tabular}
  \caption{Numerical values of the model parameters.}\label{table1}
\end{table}

\section{Numerical results}
\label{sec:2}
\subsection{Characterization of the entangled state}
\label{subsec:2a}
First, we have used the numerical model to characterize the dimensionality of the entangled bi-photon state in space and time at the output of the crystal. From time-integrated mean intensities in the near-field and in the far-field and space-integrated mean time and wavelength spectra of the signal and idler beams, we have calculated the widths of SPDC beams in space and time domains, expressed in standard deviations, which gives : 

\begin{eqnarray}\label{eq1}
\left\{\begin{array}{c}
\sigma^{SPDC}_x\simeq\sigma^{SPDC}_y\simeq 5.2\times10^{-2}\,mm\\
\sigma^{SPDC}_{\nu_{x}}\simeq\sigma^{SPDC}_{\nu_{x}}\simeq 38.2\,mm^{-1}\\
\sigma^{SPDC}_{t}\simeq 40.3\,ps\\
\sigma^{SPDC}_{\lambda}\simeq 0.14\,nm\\ 
\end{array}\right.
\end{eqnarray}   

\begin{figure}
\centering
\includegraphics[width=15cm]{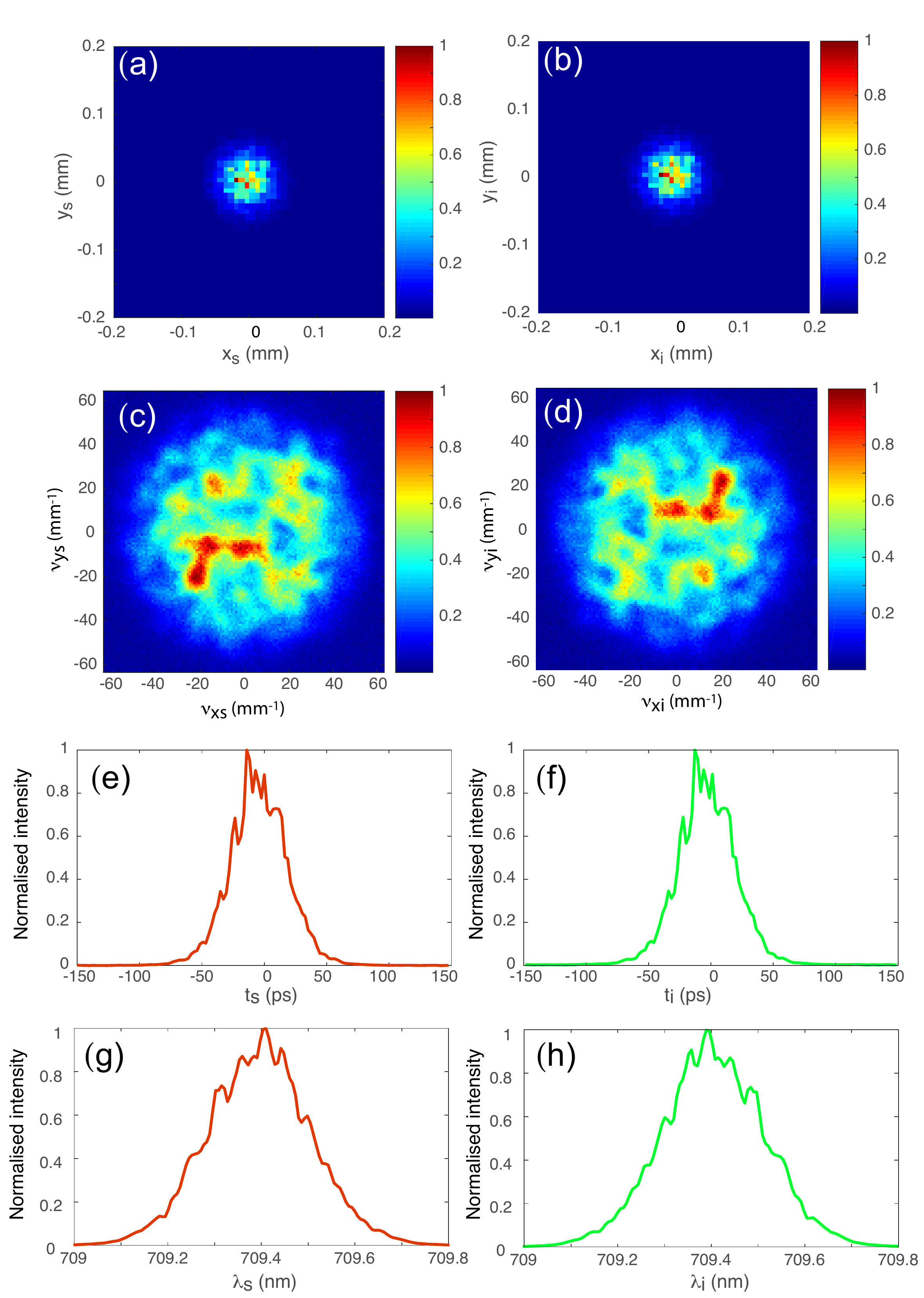}
\caption{For a single numerical realization, normalized intensity in the near-field (a,b) and in the far-field (c,d), in time (e,f) and the wavelength spectra (g,h) of the signal and idler beams, respectively.}\label{fig3}
\end{figure}

In order to evidence the correlations in space and time, we calculate the SPDC intensities obtained with a single numerical realization (Fig. \ref{fig3}) when the amplification gain is significantly increased. In good agreement with the correlation properties in space and time domains of SPDC, we can observe in near-field domain and in the temporal domain correlated fluctuations between pixels of transverse coordinates $\boldsymbol{r_s}$ and $\boldsymbol{r_i}=\boldsymbol{r_s}+\Delta\boldsymbol{r}$ (Fig.\ref{fig3} a,b) and between times $t_s$ and $t_i=t_s+\Delta t$ (Fig.\ref{fig3} e,f). On the other hand, in the far-field domain, correlated fluctuations are observed between pixels of opposite coordinates (Fig. \ref{fig3} c,d), corresponding to the opposite transverse momenta $\boldsymbol{q_s}$ and $\boldsymbol{q_i}= -\boldsymbol{q_s}+\Delta\boldsymbol{q}$. Momentum coordinates are related to the spatial frequencies by $q_{x,y}=2\pi\nu_{x,y}$. In the wavelength domain, correlations are observed between the conjugated wavelengths $\lambda_s$ and $\lambda_i=(\lambda^{-1}_{p}-\lambda^{-1}_{s})^{-1} +\Delta\lambda$ (Fig. \ref{fig3}g,h). $\Delta\boldsymbol{r}$, $\Delta\boldsymbol{q}$, $\Delta t$ and $\Delta\lambda$ denote the correlation uncertainties in the different domains that can be associated to the coherence properties of the bi-photon state. 

\begin{figure}
\centering
\includegraphics[width=15cm]{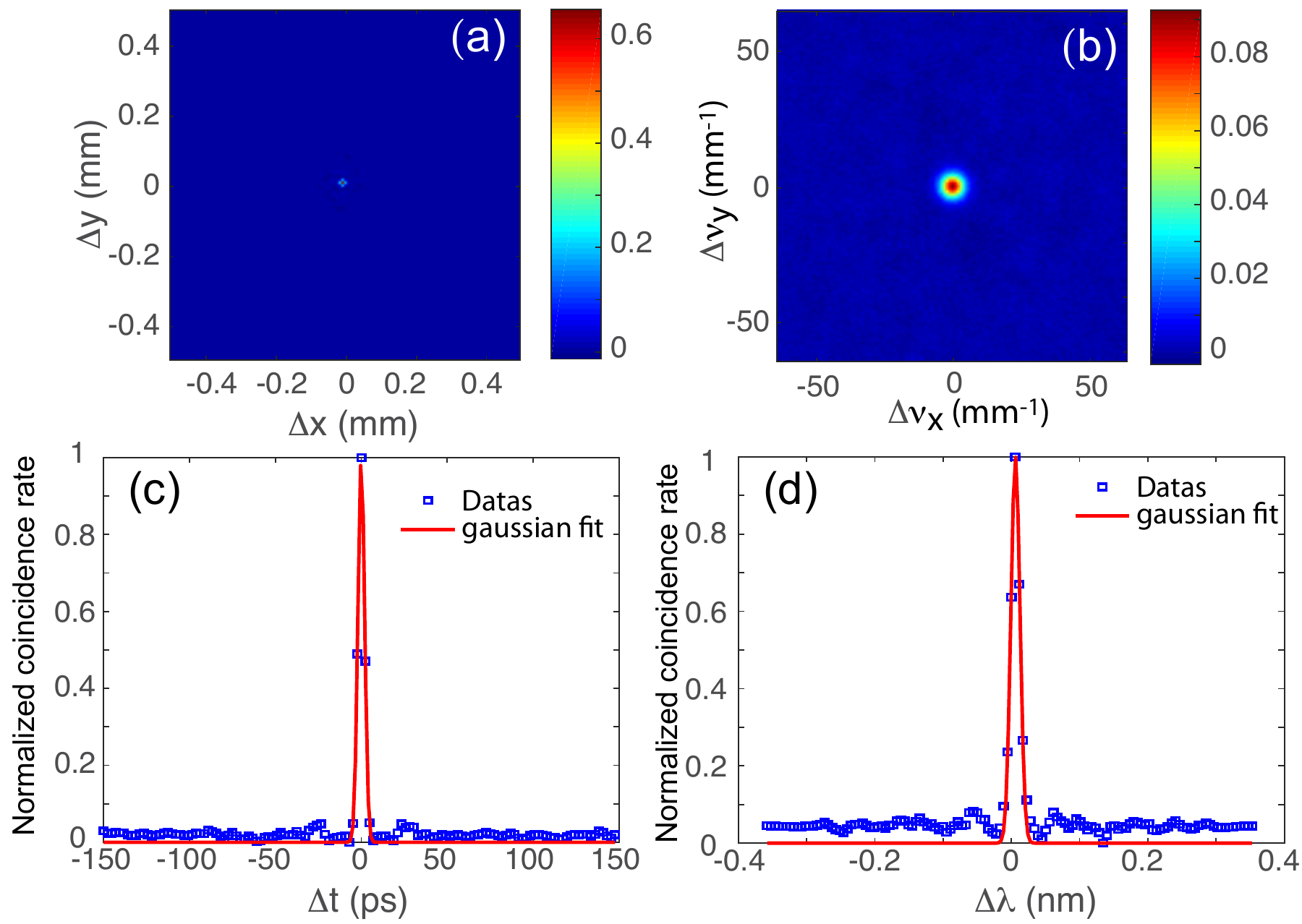}
\caption{Normalized coincidence counts rates (a) in the near-field, (b) in the far-field, (c) in time and (d) in the wavelength domains. In (c) and (d), red curves correspond to a Gaussian fit of the numerical data depicted with blue squares.}\label{fig4}
\end{figure}

Then, from numerical results we have estimated the total Schmidt number of the bi-photon state corresponding to its whole dimensionality in space-time dimensions. For this purpose, we have calculated all the normalized cross-correlations of the twin beams in the space domain (between near-field and far-field images) and in time domain ( between time and wavelength spectra). Fig. \ref{fig4} shows these cross-correlations in the near-field (\ref{fig4}a), in the far-field (\ref{fig4}b), in time (\ref{fig4}c) and in wavelength (\ref{fig4}d). Then, we have calculated the widths of the correlation peaks in space and time domains, expressed in standard deviations, which gives : 

\begin{eqnarray}\label{eq2}
\left\{\begin{array}{c}
\sigma_{x}\simeq 8.9\times10^{-3}\,mm;\,\sigma_{y}\simeq 7.6\times10^{-3}\,mm\\
\sigma_{\nu_{x}}\simeq 5.1\,mm^{-1};\,\sigma_{\nu_{y}}\simeq 4.5\,mm^{-1}\\
\sigma_{t}\simeq 2.7\,ps\\
\sigma_{\lambda}\simeq 9.4\times10^{-3}\,nm \Rightarrow \sigma_{\nu_{t}}\simeq 5.6\,GHz (@709.4 nm)\\ 
\end{array}\right.
\end{eqnarray} 

Though the pump beam profile and the phase matching function are symmetric, we can notice that the spatial widths of the correlation peaks are not equal in both dimensions. This is due to the walk-off in the considered type 2 three-wave-mixing interaction. Indeed, for non degenerated conjugated wavelengths, the symmetry centers of near-field and far-field  spatial correlations are shifted along the horizontal axis which enlarges the widths of the correlation peaks along this direction. Consequently, the larger is the bandwidth of the IF, the larger are the widths of the correlation peaks in the $x$ dimension. With these values we have estimated the Schmidt numbers for each dimension as follows :

\begin{eqnarray}\label{eq3}
\left\{\begin{array}{c}
K_x=\frac{1}{4\sigma^2_{x}\sigma^2_{\nu_{x}}}\simeq 121\\
K_y=\frac{1}{4\sigma^2_{y}\sigma^2_{\nu_{y}}}\simeq 213\\
K_t=\frac{1}{4\sigma^2_{t}\sigma^2_{\nu_{t}}}\simeq 1094\\
\end{array}\right.
\end{eqnarray}

In the spatial domain, these values can be compared to the definition of the Schmidt number given by \cite{law_analysis_2004}. For example along the $x$ dimension it gives : $K_x=\frac{1}{4} \left(\sigma_{x_p}2\pi\sigma^{SPDC}_{\nu_x}+\frac{1}{\sigma_{x_p}2\pi\sigma^{SPDC}_{\nu_x}} \right)^2$, where $\sigma_{x_p}$ is the pump width and $2\pi\sigma^{SPDC}_{\nu_x}$ is the phase matching function width in momentum space. With this definition, a Schmidt number $K_{x}=71$ is obtained, in rather good agreement with the value given by Eq. \ref{eq3}. Finally, the full space-time dimensionality of the entangled state can be estimated as $V=\sqrt{K_xK_yK_t}\simeq 5310$. This value confirms the high dimensionality of the simulated bi-photon state.

\subsection{Simulation of the spatio-temporal Hong-Ou-Mandel dip}
\label{subsec:2b}
\begin{figure}
\centering
\includegraphics[width=15cm]{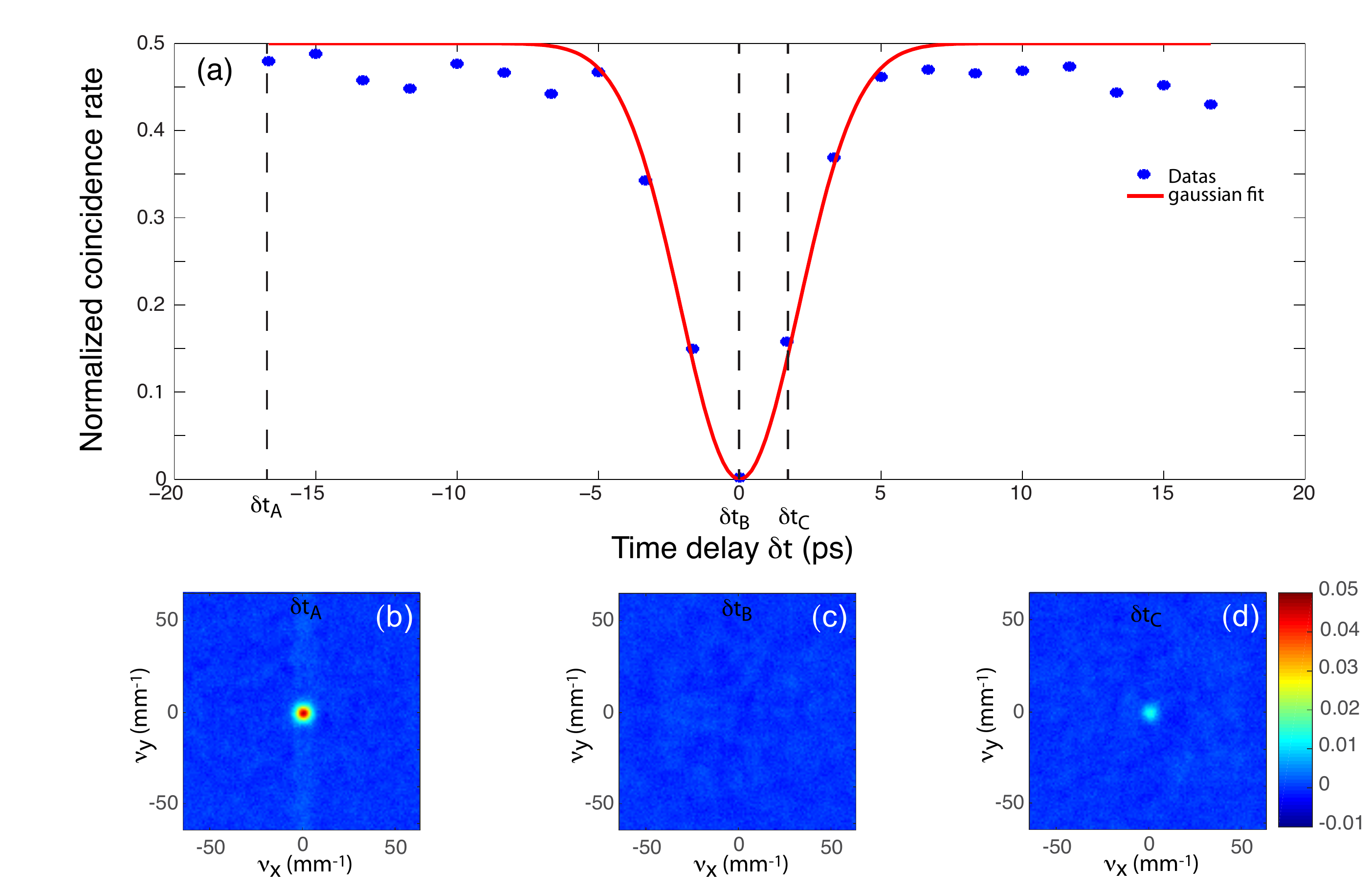}
\caption{(a) Temporal HOM dip as a function of the time delay between the signal and idler beams. The relative coincidence counts rate is obtained by summing all the spatial coincidences between the far field images of the two output ports of the BS divided by the sum of all the spatial coincidences between the far field of signal and idler images (Fig. \ref{fig4}b). The red curve corresponds to a Gaussian fit of the HOM dip and blue stars correspond to the data. The width of the temporal HOM dip, expressed in standard deviations, is $\sigma^{HOM}_t=2.95\,ps$.  (b, c, d) far field spatial coincidences between the two output ports of the BS at different time delays: $\delta t_A=-16.7\,ps$, $\delta t_B=0\,ps$, $\delta t_C=+1.7\,ps$.}\label{fig5}
\end{figure}

To simulate the two-photon interference, we have calculated the propagation of the SPDC beams from the crystal output up to the 50/50 BS, the coherent superposition of the two beams at the input ports of the BS and the time integrated detection of the far-field images at the two output ports of the BS in agreement with the setup described in Fig. \ref{fig1}. 

First we have investigated the influence of a time delay $\delta t$, between the propagation times of the signal and idler beams from the crystal output to the BS, on the coincidence counts between a set of 100 pairs of images. Fig. \ref{fig5}a shows the temporal HOM dip which depicts the relative total momentum coincidence counts rate between the two far-field images of the BS output ports as a function of the time delay.
The relative coincidence counts rate is obtained by summing all the spatial coincidences between the far-field images of the two output ports of the BS divided by the sum of all the spatial coincidences between the far-field images of signal and idler beams at the crystal output. In good agreement with the theory and the experiments \cite{hong_measurement_1987}, the coincidence rate falls to zero when the SPDC beams are perfectly synchronised ($\delta t=0$) and reaches a relative value of 0.5, with respect to the total coincidences in the absence of BS, when the time delay is much larger than the coherence time of the bi-photon wave-packet. From the numerical data fitted with a Gaussian shape, we have estimated the standard deviation of the temporal HOM dip to $\sigma^{HOM}_t=2.95\,ps$. This value is in good agreement with the standard deviation of the temporal correlation peak calculated between the signal and the idler time spectra : $\sigma_{t}\simeq 2.7\,ps$ (Fig. \ref{fig4}c). It also demonstrates that temporal properties of two-photon interference can be revealed with detectors arrays that have no temporal resolution. Figures \ref{fig5}b-d show the spatial distribution of the momentum correlation counts for different time delays. We can observe that, while the spatial width of the correlation peak remain constant, its amplitude decreases when the time delay tends toward zero. Quantitatively, in the far-field domain, the time coincidence counts rate can be expressed in momentum space as :

\begin{equation}\label{eq4}
R(\Delta\boldsymbol{q},\delta t)=\dfrac{R_0(\Delta\boldsymbol{q})}{2}\left(1-e^{-\frac{\delta t^2}{\sigma^2_{t}}}\right),
\end{equation} 
where $R_0(\Delta\boldsymbol{q})=|\Phi(\Delta\boldsymbol{q})|^2$ is the spatial distribution of the coincidence counts between the input twin beams. It is related to the far-field bi-photon wave function $\Phi(\boldsymbol{q}_s,\boldsymbol{q}_i)$ given by \cite{law_analysis_2004}:
\begin{equation}\label{eq5}
\Phi(\boldsymbol{q}_s,\boldsymbol{q}_i)=\Phi_0 e^{-\frac{|\boldsymbol{q}_s+\boldsymbol{q}_i|^2}{2 \sigma_q^2}}e^{-\frac{|\boldsymbol{q}_s-\boldsymbol{q}_i|^2}{2 \sigma_{SPDC}^2}},
\end{equation}
where $\sigma_q$ and $\sigma_{SPDC}$ are the standard deviations of the Fourier transform of the pump field and of the phase matching range in momentum units, respectively. 

\begin{figure}
\centering
\includegraphics[width=15cm]{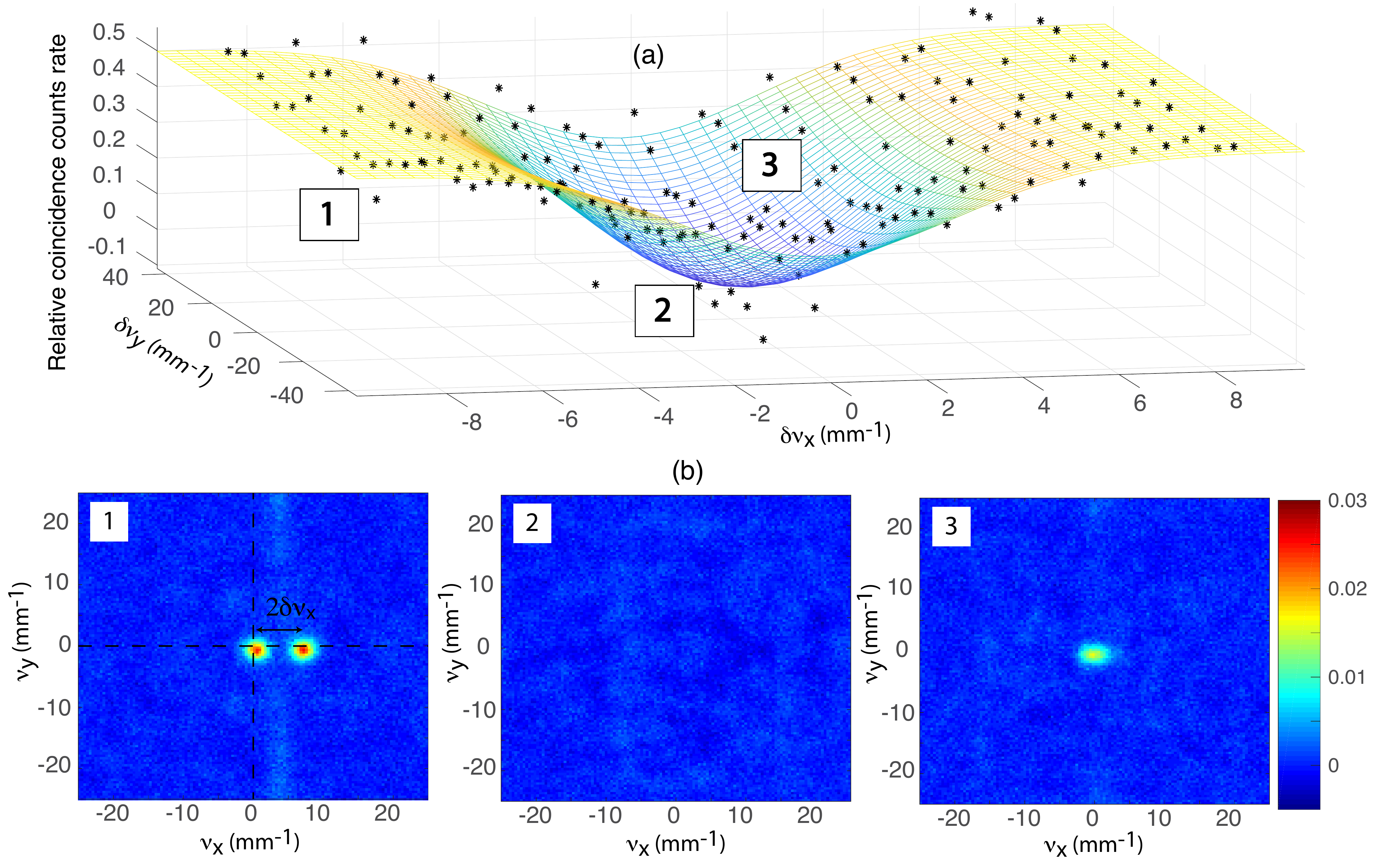}
\caption{(a) Spatial HOM dip as a function of the frequency spatial shift induced by horizontal and vertical tilts of the BS. Data are represented by black stars and the mesh surface corresponds to a Gaussian fit of data. Each point corresponds to the sum of the spatially resolved coincidence counts. (b) far-field 2D-spatial coincidence counts rate between the two output ports of the BS for different phase shifts : (1) $\delta\nu_{x_1}=-8.5\,mm^{-1}$, $\delta\nu_{y_1}=-42\,mm^{-1}$, (2) $\delta\nu_{x_2}=\delta\nu_{y_2}=0$ (3) $\delta\nu_{x_3}=1.5\,mm^{-1}$, $\delta\nu_{y_3}=28\,mm^{-1}$. In (1) the arrow denotes the relationship between positions of the two correlations peaks and the phase shift.}\label{fig6}
\end{figure}

In a second time, we have investigated the influence of a phase shift of the reflected beams, along the two transverse spatial dimensions, induced by angular tilts $\theta_{BS}$ and $\phi_{BS}$ of the BS (Fig.\ref{fig1}). The coincidence counts rate is still calculated in the far-field domain between pixels corresponding to opposite momenta. For these numerical simulations, we assume that images of the near-field of the crystal are formed at the BS interface (i.e. $d_s=d_i=0$). Moreover, no time delay ($\Delta t=0$) is considered here.

Fig. \ref{fig6}a presents, in spatial frequency units, the 2D shape of the spatial HOM dip as a function of the spatial frequency shifts along the both dimensions which are related to the angular tilts as follows : $\delta\nu_x=\dfrac{\theta_{BS}}{\lambda_p}$, $\delta\nu_y=\dfrac{\phi_{BS}}{\lambda_p}$. The black stars correspond to the numerical data and the mesh surface corresponds to a Gaussian fit of the HOM dip.
Figures \ref{fig6}b to c show the spatially resolved coincidence counts between the two outgoing images of the interferometer for different phase shifts. In Fig. \ref{fig6}b two correlation peaks can be observed : one corresponds to coincidences between the two transmitted beams and the second one is related to coincidences between the two reflected beams. The amplitudes of the correlation peaks are normalized with the amplitude of the correlation peak between the signal and the idler beams with no BS (Fig. \ref{fig4}b). In Fig. \ref{fig6}b we have drawn an arrow emphasizing the relationship between the phase shift and the relative positions of the correlations peaks. 
While the correlation peak corresponding to coincidences between both transmitted beams is centered at the zero spatial frequency, the correlation peak corresponding to coincidences between both reflected beams is shifted along the horizontal direction with an amplitude twice of the phase shift $\delta\nu_x$. No shift along the vertical direction is observed whatever the vertical tilt of the BS. Indeed, while a horizontal tilt of the BS induces a spatial frequency shift of reflected images in the same directions, a vertical tilt of the BS induces a shift of reflected images in the opposite directions. In addition to the shift of one of the correlation peaks, we can observe that the amplitude of both correlation peaks decrease when the phase shift decreases (Fig. \ref{fig6}c). Finally, when there is no phase shift, no more coincidence counts are recorded between images (Fig. \ref{fig6}b) leading to a two-photon interference visibility of 100 \%.  This result proves that twin photons exit in pairs with an equal probability through output channels 1 or 2 of the interferometer.  

With a Gaussian fit, we have estimated the standard deviations of the 2D spatial HOM dip along the horizontal dimension as  $\sigma^{HOM}_{\nu_x}=5.1\,mm^{-1}$, in good agreement with the dimension $\sigma_{\nu_x}$ of the bi-photon wave-packet (Eq. \ref{eq2}). However, along the vertical dimension the standard deviation of the HOM dip is $\sigma^{HOM}_{\nu_y}=39\,mm^{-1}$, which is much larger than the estimated standard deviation of the bi-photon wave-packet along this dimension ($\sigma_{\nu_y}=4.5\,mm^{-1}$) and corresponds to the standard deviation of phase matching (see Eq. \ref{eq1}).

In order to explain these non intuitive numerical results, we use the same formalism proposed in \cite{hong_measurement_1987} to write the analytic expression of the joint probability of detection of photons at both detectors arrays $CCD_1$ and $CCD_2$ between momenta $\boldsymbol{q}_{1}$ and $\boldsymbol{q}_{2}=-\boldsymbol{q}_{1}+\Delta\boldsymbol{q}$.  

If we take into account the momentum shifts $\delta\boldsymbol{q}_{s}$ and $\delta\boldsymbol{q}_{i}$ induced by vertical and horizontal tilts of the BS, we can write the positive-frequency parts of the fields at detectors arrays $CCD_1$ and $CCD_2$ as follows:

\begin{eqnarray}\label{eq6}
\left\{
\begin{array}{c}
\widehat{E}^{(+)}_1(\boldsymbol{q}_1)=\dfrac{1}{\sqrt{2}}\left[\widehat{E}^{(+)}_i(\boldsymbol{q}_{i,t})+i \widehat{E}^{(+)}_s(\boldsymbol{q}_{s,r}+\delta\boldsymbol{q}_s)\right]\\
\widehat{E}^{(+)}_2(\boldsymbol{q}_2)=\dfrac{1}{\sqrt{2}}\left[\widehat{E}^{(+)}_s(\boldsymbol{q}_{s,t})+i \widehat{E}^{(+)}_i(\boldsymbol{q}_{i,r}+\delta\boldsymbol{q}_i)\right]\\
\end{array}\right.
\end{eqnarray}

where $\boldsymbol{q}_{i,t}$ and $\boldsymbol{q}_{s,t}$ are the momenta of the transmitted beams. $\boldsymbol{q}_{i,r}+\delta\boldsymbol{q}_i$ and $\boldsymbol{q}_{s,r}+\delta\boldsymbol{q}_s$ are the momenta of the reflected beams. Because we assume that SPDC beams propagate in the horizontal plane, the BS induces a left-right symmetry between the transmitted and the reflected beams in the propagation plane. Then, $q_{xs,r}=-q_{xs,t}$, $q_{ys,r}=q_{ys,t}$ and $q_{xi,r}=-q_{xi,t}$, $q_{yi,r}=q_{yi,t}$. Moreover, phase shifts for both reflected images induced by a BS tilt verifies : $\delta q_{xs}=\delta q_{xi}$ and $\delta q_{ys}=-\delta q_{yi}$.
  
The joint probability of the detection of photons at both detectors arrays $CCD_1$ and $CCD_2$ at momenta $\boldsymbol{q}_{1}$ and $\boldsymbol{q}_{2}=-\boldsymbol{q}_{1}+\Delta\boldsymbol{q}$, respectively, is given by: 

\begin{equation}\label{eq7}
P_{12}\propto \langle\widehat{E}^{(-)}_1(\boldsymbol{q}_1)\widehat{E}^{(-)}_2(-\boldsymbol{q}_1+\Delta\boldsymbol{q})\widehat{E}^{(+)}_2(-\boldsymbol{q}_1+\Delta\boldsymbol{q})\widehat{E}^{(+)}_1(\boldsymbol{q}_1)\rangle.
\end{equation} 

Using equations \ref{eq6} and \ref{eq7}, we show that the joint probability is : 
\begin{equation}\label{eq8}
P_{12}\propto\dfrac{1}{2}\left(|\Phi(\boldsymbol{q}_{s,t},\boldsymbol{q}_{i,t})|^2+|\Phi(\boldsymbol{q}_{s,r},\boldsymbol{q}_{i,r})|^2 -[\Phi^*(\boldsymbol{q}_{s,t},\boldsymbol{q}_{i,t})\Phi(\boldsymbol{q}_{s,r},\boldsymbol{q}_{i,r})+cc.]\right),
\end{equation} 
where $\Phi(\boldsymbol{q}_s,\boldsymbol{q}_i)$ is the bi-photon wave function in the far field domain (see Eq. \ref{eq5}) with $\boldsymbol{q}_{s,t}=-\boldsymbol{q}_{i,t}+\Delta\boldsymbol{q}$ for the transmitted beams and $\boldsymbol{q}_{s,r}+\delta\boldsymbol{q}_s=-(\boldsymbol{q}_{i,r}+\delta\boldsymbol{q}_i)+\Delta\boldsymbol{q}$ for the reflected beams. This equation corresponds to the spatially resolved coincidence counts rate depicted by Figures \ref{fig6}b-d. The first term is related to the correlation peak centered at the zero spatial frequencies, the second one corresponds to the shifted correlation peak and the last term leads to the two-photon interference. 

Then, with Eq. \ref{eq5}, Eq. \ref{eq8} yields:
\begin{equation}\label{eq9}
P_{12}\propto\dfrac{1}{2}\left(e^{-\frac{|\Delta\boldsymbol{q}|^2}{2 \sigma_q^2}}e^{-\frac{|2\boldsymbol{q}_s-\Delta\boldsymbol{q}|^2}{2 \sigma_{SPDC}^2}}-e^{-\frac{|\Delta\boldsymbol{q}-\delta\boldsymbol{q}_{s}-\delta\boldsymbol{q}_{i}|^2}{2 \sigma_q^2}}e^{-\frac{|2\boldsymbol{q}_s-\Delta\boldsymbol{q}+\delta\boldsymbol{q}_{s}-\delta\boldsymbol{q}_{i}|^2}{2 \sigma_{SPDC}^2}} \right)^2,
\end{equation}   
 2D space integration of Eq. \ref{eq9} leads to the total coincidence counts rate as a function of the phase shift $\delta\boldsymbol{q}$ as follows: 
\begin{equation}\label{eq10}
C(\delta\boldsymbol{q})=\dfrac{C_0}{2}\left(1-e^{-\frac{\delta q^2_{x}}{\sigma_q^2}}e^{-\frac{\delta q^2_{y}}{\sigma_{SPDC}^2}}\right).
\end{equation}

In good agreement with numerical results depicted by Fig. \ref{fig6}a, this equation confirms that the spatial widths of the HOM dip depends of coherence width of the bi-photon wave-packet along the horizontal dimension and of the phase matching bandwidth along the vertical dimension. In agreement with Lee et al. \cite{lee_spatial_2006}, this results show that the observed spatial coherence is clearly linked to the geometry of the interferometer and more specifically to the odd or even number of reflections experienced by the SPDC beams.

\begin{figure}
\centering
\includegraphics[width=15cm]{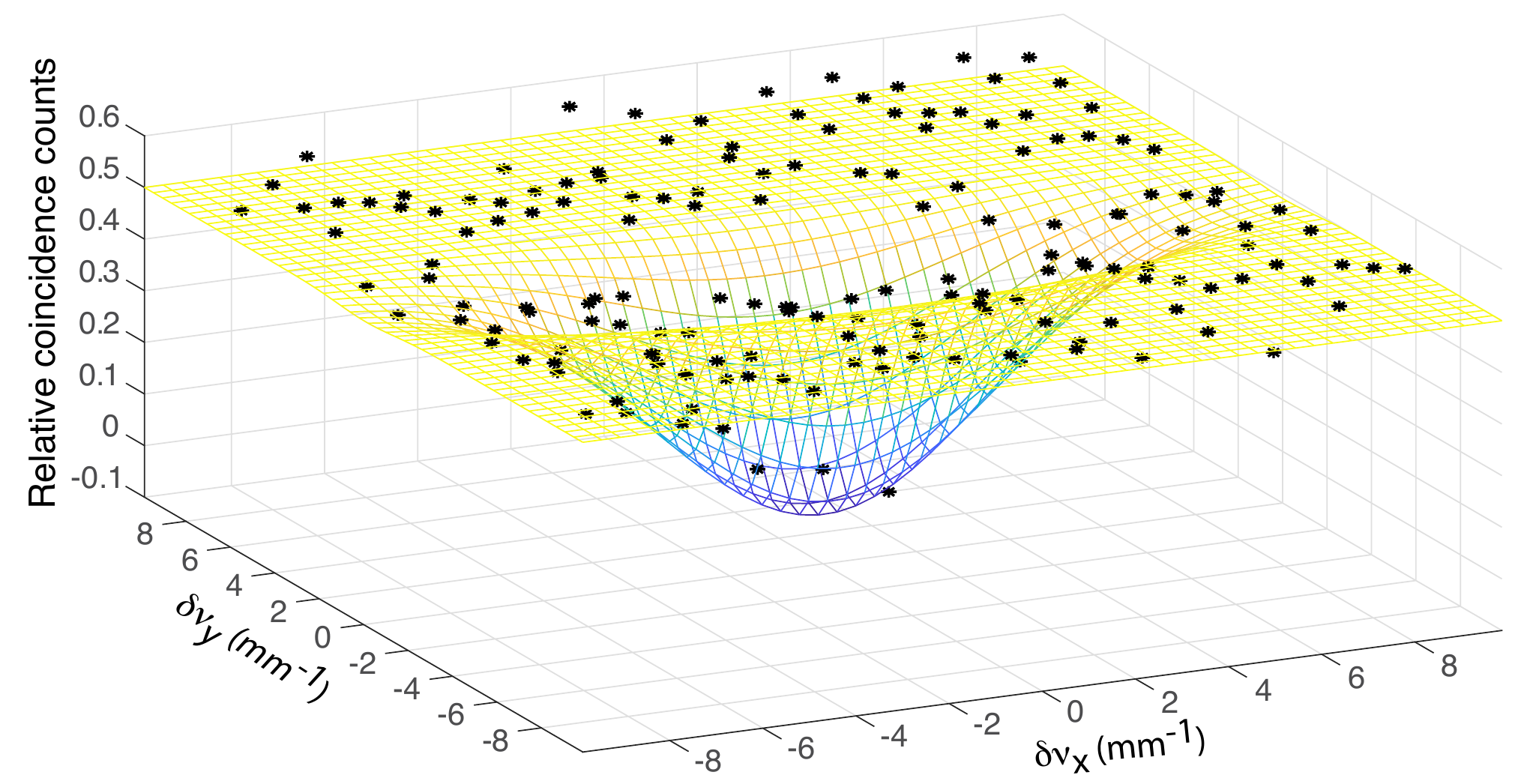}
\caption{For a 5 mm defocusing of the signal and idler image planes with respect to the BS, spatial HOM dip as a function of the frequency spatial shift induced by horizontal and vertical tilts of the BS. Data are represented by black stars and the mesh surface corresponds to a Gaussian fit of data.}\label{fig7}
\end{figure}

We have also investigated the impact of defocusing of the signal and idler image planes with respect to the BS, combined with a momentum shift of the reflected beams. To do this, we assume that the defocusing distances are equal : $d_s=d_i$. When the defocusing distances are varied between 0 and 5 mm, the numerical simulations show (Fig. \ref{fig7} depicts the HOM dip obtained with a defocusing distance of 5 mm) that two photon interference is more or less insensitive to defocusing of the image planes when the BS is tilted along the horizontal axis. In that case, the width of the HOM dip along the $x$ axis remains constant and equal to the coherence width of the bi-photon in this direction : $\sigma^{HOM}_{\nu_x}=5.0\,mm^{-1}\backsimeq\sigma_{\nu_x}$. On the contrary, along the vertical direction ($y$ axis), the two-photon interference exhibits an increasing sensitivity to the defocusing when BS is tilted along the vertical axis. Indeed, while the HOM dip is much larger along $y$ axis than the coherence width of the bi-photon and is limited by the phase-matching bandwidth when there is no defocusing, it becomes quickly narrower than the coherence width of the bi-photon when defocusing occurs : $\sigma^{HOM}_{\nu_y}=1.8\,mm^{-1}<\sigma_{\nu_y}$ for $d_s=d_i=5\,mm$.           
In this case, the coincidence counts rate given by Eq. \ref{eq10} is no more valid. A formalism involving for each beam (reflected and transmitted beams) two point spread functions (one from the image plane to the BS and one from the BS to the camera) could be developed \cite{abouraddy_entangled-photon_2002}, leading to double integrals which must be calculated for each couple of momenta ($\boldsymbol{q_1}, \boldsymbol{q_2}$). This results is an extreme difficulty to obtain an analytical solution and justifies the use of a numerical model to study the influence of experimental parameters in a realistic way when the modelled situations no longer allow the use of a simplified analytical model.  

\section{Conclusion}
\label{conclusion}
In conclusion, a numerical model based on stochastic simulations has been proposed to model the spatio-temporal properties of the two-photon interference in a typical Hong-Ou-Mandel interferometer.
With this numerical model we have properly quantified the dimensionnality of the bi-photon wave-packet in space and time and we have modeled the two-photon interference signature revealed by the calculation of the 2D spatial distribution of quantum correlations in momentum between the far-field images at the two outgoing ports of the interferometer. By tuning some parameters, like time delay between the two arms of the interferometer, spatial frequency shift of the reflected beams and defocusing of the image planes, we have fully characterized the spatio-temporal properties of the HOM dip which is related to the coherence properties of the bi-photon wave-packet. We would like to emphasize that the temporal coherence properties are obtained by the detection of purely spatial coincidences with detectors arrays that have no temporal resolution.

This numerical model is now available to simulate all the possible configurations of two-photon interference that can be proposed as well as all unitary operations that can be applied to images at the ongoing and/or at the outgoing ports of the interferometer or spatio-temporal phase shaping of the pump pulse \cite{walborn_multimode_2003}, allowing full bi-dimensional manipulation of bi-photon states with high Schmidt number.     
This numerical model is also useful for the experimental implementation of this HOM interferometer currently in progress. Indeed, it demonstrates, in good agreement with previous works \cite{lee_spatial_2006,kim_spatial_2006}, that the geometry of the interferometer is a crucial element for the measurement of two-photon interference. It also makes it possible to evaluate the sensitivity of the interference phenomenon according to the experimental parameters and the precision of the settings required to observe the HOM dip.


\end{document}